\documentclass[pre,twocolumn,showpacs]{revtex4}
\usepackage{amsmath,amssymb}
\usepackage{graphics,color}
\usepackage{times}
\usepackage{graphicx}

\newcommand{\ket}[1]{| #1 \rangle}
\newcommand{\aver}[1]{\left \langle #1 \right \rangle}

\begin{document}

\title{Quasimodes of a chaotic elastic cavity with increasing local losses}
\author{O. Xeridat}
\author{C. Poli}
\author{O. Legrand}
\author{F. Mortessagne}
\author{P. Sebbah}
\email[] {sebbah@unice.fr}
\affiliation{Universit\'e de Nice-Sophia Antipolis, Laboratoire de Physique de la Mati\`ere Condens\'ee, CNRS UMR 6622, 
Parc Valrose, 06108 Nice Cedex 02, France}

\begin{abstract}
We report non-invasive measurements of the complex field of elastic quasimodes of a silicon wafer with chaotic shape. The amplitude and phase spatial distribution of the flexural modes are directly obtained by Fourier transform of time measurements. We investigate the crossover from real mode to complex-valued quasimode, when absorption is progressively increased on one edge of the wafer. The complexness parameter, which characterizes the degree to which a resonance state is complex-valued, is measured for non-overlapping resonances and is found to be proportional to the non-homogeneous contribution to the line broadening of the resonance. A simple two-level model based on the effective Hamiltonian formalism supports our experimental results.
\end{abstract}


\maketitle

A closed quantum system is fully defined by a set of eigenenergies and orthogonal discrete states. When the system is coupled to the environment, e.g.  through leakage at the boundaries, mode lifetime becomes finite. Consequently, the spectral widths of the resonances broaden and are no longer isolated. While the statistical properties of the spectral widths of chaotic wave systems was systematically analyzed in the regime of isolated resonances \cite{Por56,Alt95}, in the overlapping regime \cite{Som99,Kuh08} and in the strong overlapping regime \cite{Sok89}, the effects of the leakage on the wave function statistics remain an issue \cite{Fyo03,Kuh05}. For a wave system whose closed limit displays time reversal symmetry, the eigenfunctions become complex-valued quasimodes due to the presence of currents, the standing-wave component being progressively replaced by a component traveling toward the system boundaries \cite{Pni96}. Such a complexness may reveal itself in current density \cite{Ish01,Hoh09} and long-range correlations of wave function intensity \cite{Bro03,Kim05}. Analysis of the non-real nature of the field appears also in various domains of wave physics. It is an essential ingredient in the theory of lasing modes, which induces an enhancement of the line width as pointed out by Petermann \cite{Pet79} and studied in details for  chaotic lasing cavities \cite{Sch00}. The complex nature of the field was also discussed recently in the context of disordered open media \cite{Van09} and in diffusive random lasers \cite{Tur08}. To quantify the complexness of the field, it is convenient to introduce the ratio of the variances of the imaginary and real parts of the field as a single parameter: the complexness parameter $q^2$ \cite{Lob00}. Measuring $q^2$ requires the complete knowledge of the spatial distribution of the field, including its amplitude and phase. Indirect estimation of the complexness parameter assuming ergodicity over several resonances was reported in \cite{Bar05}. But, to the best of our knowledge, direct measurements for a given quasimode has not been reported.

In this work, we measure non-invasively Lamb waves propagating on a doubly-truncated silicon wafer and measure the effect of increasing losses on spectral and spatial characteristics of the modes. Leakage channels are progressively opened by sticking absorbent strips with increasing dimensions along one edge of the chaotic sample. For each configuration, the acoustic field spatial distribution, the phase probability distribution, the complexness parameter $q^2$ and the spectral width $\Gamma$ are measured for individual non-overlapping resonances. Experiments show a simple proportionality between $q$ and $\Gamma$ when the losses are varied. Experimental results are found to be consistent with a simple analytical 2-level model based on the scattering approach of open systems.

The complexness parameter of the $n$th quasimode $\Psi_n$ is defined by: 
\begin{equation}
\label{q2def}
q^2_n=\frac{\aver{\textrm{Im}(\Psi_n)^2}}{\aver{\textrm{Re}(\Psi_n)^2}} \, ,
\end{equation}
where the triangular brackets denote the spatial average.  It is worth noting that $q^2$ is equal to 0 for a closed cavity with no currents and tends to 1 for a pure traveling-wave in open space.  
 
The derivation of the probability distributions of the complexness parameter in the perturbative regime was derived in \cite{Pol09b} using the effective Hamiltonian formalism and applying a random matrix approach to open systems (see \cite{Dit00,Oko03} for reviews). While a  $N$-level model with $N\rightarrow \infty$ is needed for the probability distribution, a 2-level model \cite{Sav06,Pol09a} is here sufficient to consider the relationship between the spectral widths and the complexness parameter. We start from the effective Hamiltonian $H_{eff} =H-iVV^T/2$, where $H$ is the Hamiltonian of the closed system modeled by a $2\times 2$ matrix  and $iVV^T/2$ is an imaginary potential describing the coupling to the environment in terms of $M$ open channels. The $2 \times M$ matrix $V$ contains the coupling amplitudes $V_{n}^c$ which couple the $n$th level to the $c$th open channel. As a result of the non-hermiticity of the effective Hamiltonian, its eigenvalues and eigenvectors are complex. The eigenenergies of $H_{eff}$ reads $\mathcal{E}_{1,2}=\epsilon_{1,2}-\frac{i}{2}\Gamma_{1,2}$, where $\epsilon_{1,2}$ and $\Gamma_{1,2}$ are, respectively, the two eigenenergies and  the two spectral widths of the 2-level model. In the eigenbasis of $H$  the effective Hamiltonian is written as
\begin{equation}\label{mod}
H_{eff} =\begin{pmatrix} 
E_1 &  & 0\\
0 &  & E_2 
\end{pmatrix} 
-\frac{i}{2} \begin{pmatrix} 
\Gamma_{11} &\Gamma_{12} \\
\Gamma_{21} & \Gamma_{22}   \\
\end{pmatrix}\, ,
\end{equation}
where $E_{1,2}$ are the eigenenergies of $H$ ($E_2>E_1$ is assumed) and $\Gamma_{np}=\sum_{c=1}^M V_{n}^cV_{p}^c$. As we focus on the isolated resonance regime, the imaginary potential may be viewed as a perturbation of the Hermitian part and the effective Hamiltonian (\ref{mod}) can be easily diagonalized through a first order perturbation theory\footnote{Note that $H$ includes a Hermitian part leading to a downward shift of the eigenenergies (as observed for the central frequencies of the modes in Fig. \ref{fig:3}). Still, in the eigenbasis of $H$, the statistical assumption concerning the imaginary part of the effective Hamiltonien is not altered.}. One obtains straightforwardly the eigenenergies $\epsilon_{1,2}=E_{1,2}$, the spectral widths $\Gamma_{1,2}=\sum_{c=1}^M (V_{1,2}^c)^2 $, while the perturbed eigenvectors $\ket{\psi_1}$ and $\ket{\psi_2}$, read, in the basis $\{ \ket{1},\ket{2} \}$ of $H$: $\ket{\psi_1}=\ket{1}-if\ket{2}$ and $\ket{\psi_2}=\ket{2}+if\ket{1}$, with $f=\Gamma_{21}/(2(E_2-E_1))=\Gamma_{12}/(2(E_2-E_1))$ (in this model $\Gamma_{12}=\Gamma_{21}$).

In the 2-level model, $q^2_n$ is obtained using (\ref{q2def}) by replacing the spatial average by an average over the components of the eigenvectors $\ket{\psi_1}$ and $\ket{\psi_2}$. It reads $q^2=\Gamma_{21}^2/(4(E_2-E_1)^2)$, for both resonances . For a uniform increase of the inhomogeneous losses \cite{Bar05a}, each coupling amplitude increases in the same way: $V\rightarrow \sqrt{v} V$, where $\sqrt{v}$ characterizes the enhancement of the losses. As a result, the spectral width and the complexness parameter depend on $v$ as: $\Gamma\to v\Gamma$ and $q^2\to v^2q^2$
implying a linear relation between $\Gamma$ and $q$ for a given mode
\begin{equation}\label{qG1}
q_n=\frac{\sqrt{(\sum_c V_1^cV_2^c)^2}}{\sum_c (V_n^c)^2}\frac{\Gamma_n}{2(E_2-E_1)}\, .
\end{equation}
The relationship can be written under the general form $q_n= \beta_n \Gamma_n$,
where the slope $\beta_n$ depends on the resonance because of the fluctuations of both coupling amplitudes and spectrum \cite{Sto99}. In the limiting case of a large number of weakly coupled channels  ($M\rightarrow \infty$, $\sigma^2\rightarrow 0$ and $\aver{\Gamma}=M\sigma^2$ fixed) the term depending on the coupling amplitudes is proportional to $1/\sqrt{M}$ as reported in \cite{Sav06}, such that the fluctuations of the slope are only due to the spectrum. In the following we will only focus on the proportionality between $q_n$ and $\Gamma_n$ considering a 2D-chaotic acoustic cavity.

Two kinds of elastic waves can propagate in thin plates: horizontally polarized shear waves (SH) and Lamb waves, which are a combination of vertically polarized shear waves (SV) and longitudinal waves (L). SV- and L-waves couple with each other on plate/air interfaces and provide symmetric and antisymmetric displacements \cite{Gra91}. The only guided modes which exist at all frequencies are the zero-order symmetric $S_0$ and antisymmetric $A_0$ modes, also called extensional and flexural modes in the low frequency limit (wavelength$\gg$plate thickness). In the low frequency range, $A_0$ is highly dispersive.
Here we consider Lamb waves propagating on 2" silicon wafers of thickness $h=380 \mu$m.
The initial silicon wafer is cut at $R/2$ to form a D-shape plate. The associated classical billiard is known to display chaotic dynamics of ray trajectories \cite{Ree99}. This chaotic shape ensures mode ergodicity as opposed to integrable cavities, such as circular ones, where modes have regular patterns. The modal statistics in such a chaotic cavity is essentially Gaussian with spectral repulsion between nearest eigenfrequencies \cite{Ber77}. Another cut perpendicular to the first one is made at $\sqrt{3}R/2$ to break the remaining symmetry. The resulting shape is shown in inset in Fig.~\ref{fig:4}.

The acoustic waves are optically excited using a frequency-doubled Nd:YAG Q-switched laser (Quantel Ultra) operating at wavelength 532 nm and 8 ns pulse width. The laser pulse heats the surface of the sample and creates a few $\mu s$-long acoustic pulse by thermoelastic effect. The beam is focused down to a 200 $\mu$m-diameter spot, which corresponds to a fluence of 0.48 J/cm$^2$. This is below ablation threshold \cite{Kor09} and is a good compromise between bandwidth and signal-to-noise ratio.

On the other side of the wafer, a \textit{Thal\`es} SH-140 interferometric heterodyne optical probe measures a time response proportional to the plate normal displacement at one point over a wide bandwidth [20 kHz-45 MHz]. The resolution of $10^{-4}$ \AA/$\sqrt{\textrm{Hz}}$ allows a sensitivity to displacements of the order of a few angstr\"oms \cite{Roy86}. As our probe is only sensitive to the normal displacement of the plate, we detect preferentially $A_0$. The time sequence of 500 thousand data points is recorded by a 400 MHz-bandwidth \textit{Lecroy WaveSurfer} digital oscilloscope triggered by the laser pulse at a repetition rate of 20 Hz, and averaged over 100 traces. The laser excitation and the optical detection provide a totally non-invasive setup to investigate the dynamics of free acoustic wave propagation.

The sample is supported horizontally by three pins. We checked that their impact on acoustic propagation is negligible. The main source of losses comes from coupling with air. To scan the entire wafer, a XY-translation stage is used with a 500 $\mu$m-spatial resolution, while the optical probe remains still. Mirrors are attached to the stages to ensure that the excitation hits the sample at the same position after each displacement.
\begin{figure}
 \centering
  \includegraphics[width=80mm]{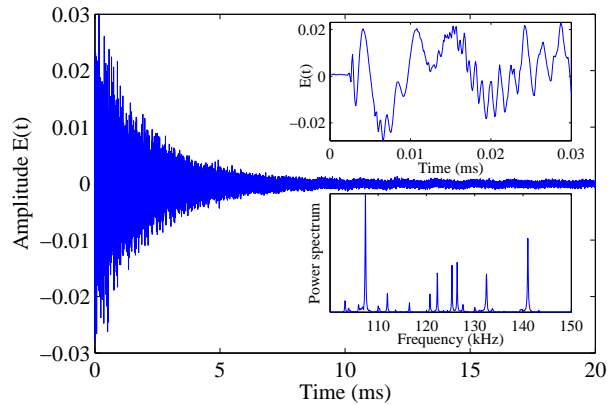}
  \caption{Time measurement at one point of the sample. Inset: (top) beginning of the signal, (bottom) corresponding spectrum.}
  \label{fig:1}
\end{figure}
Figure \ref{fig:1} shows a typical time-signal recorded at one point of the sample. The dispersive predominant antisymmetric $A_0$ mode reaches the optical detector after a few $\mu s$ and is subsequently reflected at the edges of the wafer. Neither the symmetric $S_0$ Lamb mode foregoing $A_0$, nor the SH-mode are detected by the optical probe. The entire multiply-scattered exponentially decaying signal lasts several milliseconds. Time records are stored for each value of X and Y. Hence we reconstitute the space-time map of the normal displacement field at the surface of the wafer. A speckle-like spatial distribution is rapidly reached as a result of the chaotic geometry of the doubly-truncated wafer \cite{Dra97}.

\begin{figure}
  \centering
  \includegraphics[width=75mm]{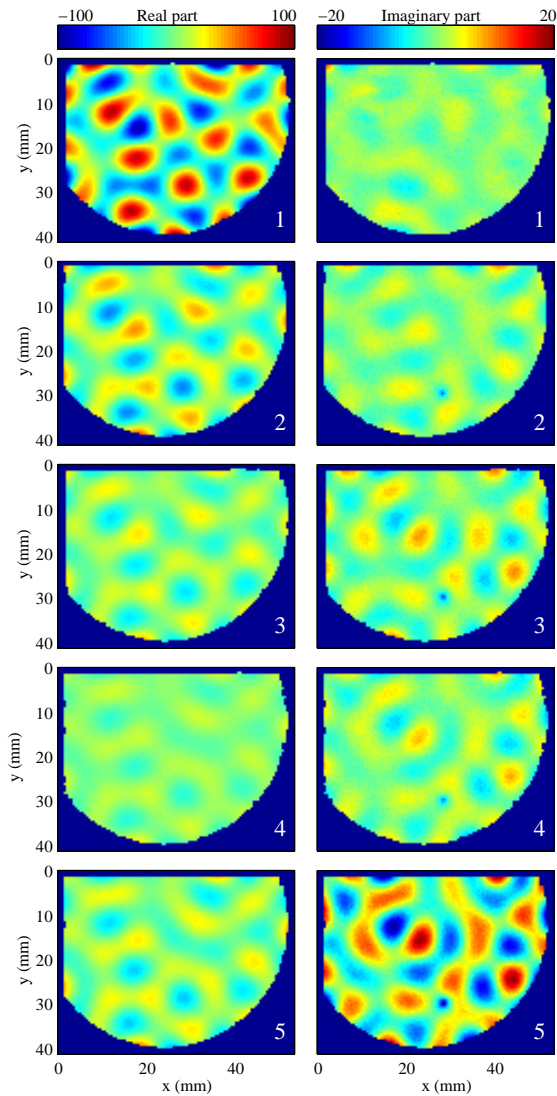}
  \caption{(Color online.) Real and imaginary component of a quasimode at its central frequency for: (1) no absorber, (2) 1 mm-thick, 45 mm-long and 1 mm-lage strip, (3) 2 mm-thick, 45 mm-long and 1 mm-broad strip, (4) same as (3) with a 2 mm-broad strip, (5) same as (3) with a 3 mm-broad strip.}
  \label{fig:2}
\end{figure}

\begin{figure}
\centering
\includegraphics[width=75mm]{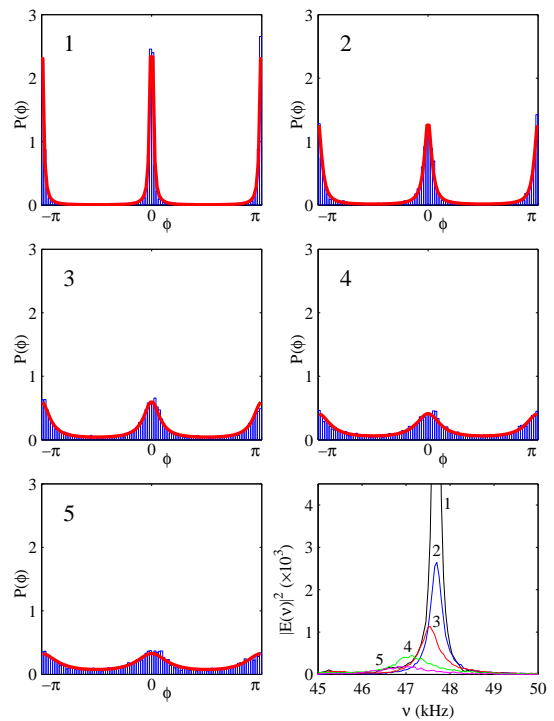}
\caption{(Color online.) Phase probability distribution of the quasimode for each of the 5 configurations described in Fig.~\ref{fig:2}, compared with Eq.~(\ref{pPhi}) with values of $q^2$ obtained from the ratio of the variances of the imaginary and real parts of the quasimode at its central frequency. (Bottom right) Corresponding spectra.}
\label{fig:3}
\end{figure}

\begin{figure}[h]
 \centering
  \includegraphics[width=75mm]{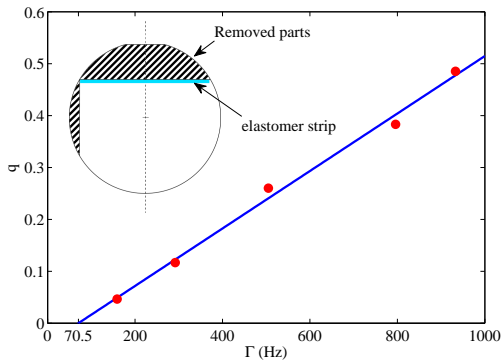}
  \caption{Square root of complexness parameter versus spectral width measured for the quasimode at its central frequency for the 5 configurations described in Fig.~\ref{fig:2} and linear fit to the data points. Inset: doubly-truncated 2"-diameter silicon wafer.}
  \label{fig:4}
\end{figure}

Each time-record associated with each point on the wafer is Fourier-transformed to obtain the spatial distributions of the real and imaginary parts of the acoustic field as a function of frequency. An example of a power spectrum measured at one point is shown in lower inset of Fig.~\ref{fig:1}. Each peak corresponds to a vibration mode of the plate. The intensity of each peak depends on the overlap of the mode with the excitation and detection. For a well isolated resonance, one obtains the maps of the real and imaginary parts of the corresponding mode. An example is shown in Fig.~\ref{fig:2}.
We measure the complexness parameter and the spectral width for individual non-overlapping modes in a set of experiments where absorption is increased locally on one edge of the sample. Here we are interested in relating $q$ to $\Gamma$, which comes down to comparing the change in the spatial nature of the mode with its spectral characteristics. This is accomplished by sticking absorbent strips with different dimensions on the wafer's longest edge, as shown in inset of Fig.~\ref{fig:4}. The absorbent is a PolyDiMethylSiloxane (PDMS) elastomer (Dow Corning, Sylgard 184). A 10:1 PDMS:cross-linker mixture was stirred and degassed then cured at $80\,^{\circ}{\rm C}$ for 2 hours. While the width and the thickness of the polymer strips are varied, the length is kept constant to fix the number of opened channels. The degree of acoustic absorption of each strip is not fully controlled and depends for instance on the way the strip is stuck on the wafer. However, the relevant parameters are measured for a given value of the absorption whatever it can be.
For each configuration, the spectrum around 47 kHz where an isolated resonance has been identified is plotted in Fig.~\ref{fig:3}. As the absorption increases, the central frequency of the resonance is shifted toward lower frequencies while its spectral width broadens. The real and imaginary parts of the quasimode are represented in Fig.~\ref{fig:2} for each sample configuration. The increasing absorption results in a global damping of the real part, whereas the imaginary contribution increases on average. This is accompanied by a progressive deformation of the field spatial distribution. 

To obtain the complexness parameter, modes with Gaussian statistics are selected, excluding for instance scarred modes \cite{Kap98}, and a phase rotation is performed on the measured field $\psi_n$: $\Psi_n=e^{i\alpha}\psi_n$ in order for the real and imaginary parts to be independent variables \cite{Ish01}. The phase $\alpha$ is unique and is fixed by the constraint $\aver{\textrm{Im}(\Psi_n)\textrm{Re}(\Psi_n)}=0$ resulting in a probability distribution of the phase $\varphi=\arctan(\textrm{Im}(\Psi)/\textrm{Re}(\Psi))$ given by \cite{Lob00}:
 \begin{equation}\label{pPhi}
p(\varphi)=\frac{q}{2\pi}\frac{1}{q^2\cos^2 \varphi+\sin^2\varphi}
\end{equation}
This expression is peaked around $0$ and $\pi$ for purely standing waves in a closed cavity. As channels are increasingly opened, the phase probability distribution broadens as seen in Fig.~\ref{fig:3}, corresponding to a growing traveling-wave component of the mode.
The linewidths are obtained by fitting the spectra with a complex Lorentzian $L(\omega)=C/(\omega-\omega_0+i\Gamma)$, where the central frequency $\omega_0$, the spectral width $\Gamma$ and the complex constant $C$ are fitting parameters. The complexness parameter is obtained from the ratio of the variances of the imaginary and real parts of the quasimode at its central frequency. Figure \ref{fig:4} shows a linear dependence of q vs. $\Gamma$. The deviation from the linear fit $q=5.54\times10^{-4}-0.04\Gamma$ is below 3\%. The complexness parameter extrapolates to 0 for $\Gamma$=70.5Hz. This value corresponds to the homogeneous contribution to the losses resulting from uniform surface-coupling with air. This losses contribute only to a uniform decay rate of the mode. Only losses localized at the sample edges contribute to the complex nature of the mode. Different slopes are found for different modes. A systematic study over a larger number of quasimodes should allow a statistical exploration of the spatial and spectral characteristics of the resonance states and should provide for an experimental check of the statistical features of the complexness parameter \cite{Pol09b}.

\begin{acknowledgments}
We thank C. Vanneste and D. Savin for fruitful discussions, O. Gauthier-Lafaye and coworkers at the RTB-LAAS plateform (UPR8001) for providing some of the samples and X. Noblin for preparing the polymer strips. This work was supported by the CNRS (PICS $\#2531$ and PEPS 07-20) and the ANR under grant ANR-08-BLAN-0302-01.
\end{acknowledgments}

\end{document}